\documentclass[12pt,preprint]{aastex}

\slugcomment{To be submitted to ApJ}

\shorttitle{Tilting Jupiter and Saturn ...}
\shortauthors{Vokrouhlick\'y \& Nesvorn\'y}

\begin{document}

\title{Tilting Jupiter (a bit) and Saturn (a lot) During \\Planetary Migration}

\author{David Vokrouhlick\'y}
\affil{Institute of Astronomy, Charles University,
       V Hole\v{s}ovi\v{c}k\'{a}ch 2, Prague 8, CZ-180 00,
       Czech Republic}
\email{vokrouhl@cesnet.cz}

\and

\author{David Nesvorn\'y}
\affil{Department of Space Studies, Southwest Research Institute,
       1050 Walnut Street, Suite 300, \\ Boulder, CO 80302, USA}
\email{davidn@boulder.swri.edu}

\begin{abstract}
 We study the effects of planetary late migration on the gas giants
 obliquities. We consider the planetary instability models from
 \citet{nm12}, in which the obliquities of Jupiter
 and Saturn can be excited when the spin-orbit resonances occur. The
 most notable resonances occur when the $s_7$ and $s_8$ frequencies,
 changing as a result of planetary migration, become commensurate with the
 precession frequencies of Jupiter's and Saturn's spin vectors. We show that
 Jupiter may have obtained its present obliquity by crossing
 of the $s_8$ resonance. This would set strict constrains on the character of
 migration during the early stage. Additional effects on Jupiter's obliquity
 are expected during the last gasp of migration when the $s_7$ resonance was
 approached. The magnitude of these effects depends on the precise value of
 the Jupiter's precession constant. Saturn's large obliquity was likely
 excited by capture into the $s_8$ resonance. This
 probably happened during the late stage of planetary migration when
 the evolution of the $s_8$ frequency was very slow, and the conditions
 for capture into the spin-orbit resonance with $s_8$ were
 satisfied. However, whether or not Saturn is in the spin-orbit resonance with
 $s_8$ at the present time is not clear, because the existing observations of
 Saturn's spin precession and internal structure models have significant
 uncertainties.
\end{abstract}


\section{Introduction}
It is believed that the orbital architecture of the solar system was
significantly altered from its initial state after the dissipation of
the protosolar nebula. The present architecture is probably a result of
complex dynamical interaction between planets, and between planets and
planetesimals left behind by planet formation. This becomes apparent
because much of what we see in the solar system today can be explained if
planets radially migrated, and/or if they evolved through a dynamical
instability and reconfigured to a new state \cite[e.g.,][]{m95,tetal99,tetal05}.

While details of this process are not known exactly, much has been learned
about it over decade by testing various migration/instability models against
various constraints. Some of the most important constraints are provided by
the terrestrial planets and the populations of small bodies in the asteroid and
Kuiper belts \cite[e.g.,][]{getal05,mm09,mm11,letal08,metal10,n15}. Processes
related to the giant planet instability/migration were also used to explain
capture and orbital distribution of Jupiter and Neptune Trojans
\cite[e.g.,][]{metal05,nv09,netal14a} and irregular satellites
\cite[e.g.,][]{netal07,netal14b}. Some of the most successful instability/migration
models developed so far postulate that the outer solar system contained
additional ice giant that was ejected into interstellar space by Jupiter
\cite[e.g.,][]{n11,nm12,betal12}. The orbits of the four surviving giant
planets evolved in this model by planetesimal-driven migration and by
scattering encounters with the ejected planet. In this work, we use this
framework to investigate the behavior of Jupiter's and Saturn's obliquities.

The obliquity, $\varepsilon$, is the angle between the spin axis of a planet
and the normal to its orbital plane. The core accretion theory applied to
Jupiter and Saturn implies that their primordial obliquities should be very
small. This is because the angular momentum of the rotation of these planets
is contained almost entirely in their massive hydrogen and helium envelopes.
The stochastic accretion of solid cores should therefore be irrelevant for their
current obliquity values (see Lissauer \& Safronov 1991 for a discussion),
and a symmetric inflow of gas on forming planets should lead to $\varepsilon
\simeq 0$. The present obliquity of Jupiter is $\varepsilon_{\rm J}=3.1^\circ$,
which is small, but not quite small enough for these expectations,
but that of Saturn is $\varepsilon_{\rm S}=26.7^\circ$, which is clearly not.

\citet{wh04} and \citet{hw04} noted that the precession frequency of Saturn's
spin axis has a value close to $s_8 \simeq -0.692$ arcsec yr$^{-1}$, where $s_8$
is the mean nodal regression of Neptune's orbit (or, equivalently, the eighth
nodal eigenfrequency of the planetary system; e.g., Applegate et~al. 1986;
Laskar 1988). Similarly, \citet{wc06} pointed out that the precession frequency
of Jupiter's spin axis has a value close to $s_7 \simeq -2.985$ arcsec yr$^{-1}$,
where $s_7$ is the mean nodal regression of Uranus's orbit. These findings are
significant because they raise the  possibility that the current obliquities
of Jupiter and Saturn have something to do with the precession of the giant
planet orbits. Specifically, \citet{wh04} and \citet{hw04} suggested that the
present value of Saturn's obliquity can be explained by capture of Saturn's
spin vector in a resonance with $s_8$. They proposed that the capture occurred
when Saturn's spin vector precession increased as a result of Saturn's cooling
and contraction, or because $s_8$ decreased during the depletion of the
primordial Kuiper belt. They showed that, if the post-capture evolution is
conveniently slow, the spin-orbit resonance (also known as the Cassini
resonance, see Section 2) is capable of increasing Saturn's obliquity to its
current value.

While changes of precession or $s_8$ during the earliest epochs could have
been important, it seems more likely that capture in the spin-orbit
resonance occurred later, probably as a result of planetary migration
\citep{betal09}. This is because both $s_7$ and $s_8$ significantly change
during the instability and subsequent migration. Therefore, if the spin-orbit
resonances had been established earlier, they would not survive to the present
time. \citet{betal09} studied various scenarios for resonant tilting of Saturn's
spin axis during the planetary migration and found that the present obliquity
of Saturn can be explained only if the characteristic migration time scale was
long and/or if Neptune's inclination was substantially excited during the
instability. Since Neptune inclination is never large in the instability/migration
models of Nesvorn\'y \& Morbidelli (2012; hereafter NM12), typically $i_{\rm N}<
1^\circ$, the migration timescales presumably need (note that \citet{betal09} did
not investigated these low-$i$ cases in detail) to be {\it very} long (see
Section~3.3). Interestingly, these very long migration timescales are also
required from other constraints \cite[e.g.,][]{metal14,n15}.
They could be achieved in the \citet{nm12} models if Neptune interacted
with an already depleted planetesimal disk during the very last stages of the
migration process. As for the obliquity of Jupiter, \citet{wc06} suggested that
the present value is due to the proximity of the spin precession rate to
the $s_7$ frequency.

In fact, the obliquities of Jupiter and Saturn represent a much stronger
constraint on the instability/migration models than was realized before. This
is because the constraints from the present obliquities of Jupiter and Saturn
must be satisfied {\it simultaneously} (McNeil \& Lee 2010; Brasser \& Lee 2015).
For example, in the initial configuration of planets in the NM12 models, the
$s_8$ frequency is much faster than the precession constants of both Jupiter
and Saturn. This means that the $s_8$ mode, before approaching Saturn's
precession frequency and exciting its obliquity, must also have evolved
over the precession frequency of Jupiter's spin vector. This leads to a
conundrum, because if the crossing were slow, Jupiter's obliquity would
increase as a result of capture into the spin-orbit resonance with $s_8$. If,
on the other hand, the general evolution at all stages were fast, the conditions
for capture of Saturn into the spin-orbit resonance with $s_8$ may not be be met
\cite[e.g.,][]{betal09}, and Saturn's obliquity would stay small.%
\footnote{Note that the present obliquities of Uranus and Neptune are not an
 important constraint on planetary migration, because their spin precession
 rates are much slower than any secular eigenfrequencies of orbits (now or in
 the past). Therefore, the secular spin-orbit resonances should not occur for
 these planets, and giant impacts may be required to explain their obliquities
 \cite[e.g.,][and references therein]{metal12}.}

A potential solution of this problem would be to invoke fast evolution of
$s_8$ early on, during the crossing of Jupiter's precession frequency, and
slow evolution of $s_8$ later on, such that Saturn's spin vector can be
captured into the spin-orbit resonance with $s_8$ during the late stages. This
can be achieved, for example, if the migration of the outer planets was faster
before the instability, and slowed down later, as the outer Solar System progressed
toward a more relaxed state. As we show in Section 3.1, the jumping-Jupiter models
developed in NM12 provide a natural quantitative framework to study this possibility.

In Sec.~\ref{meth} we first briefly review the general equations for the
spin-orbit dynamics. Then, in Sec.~\ref{res}, we investigate the behavior of the
spin vectors of Jupiter and Saturn in the instability/migration models of NM12.
We find that the constraints posed by Jupiter's and Saturn's obliquities can be
satisfied simultaneously in this class of models, and derive detailed conditions
on the migration timescales and precession constants that would provide a
consistent solution.

\section{Methods} \label{meth}

\subsection{Parametrization using non-singular spin vector components}
Consider a planet revolving about the Sun and rotating with angular
velocity $\omega$ about an instantaneous spin axis characterized by
a unit vector ${\bf s}$. With solar gravitational torques applied
on the planet, $\omega$ remains constant, but ${\bf s}$
evolves in the inertial space according to \cite[e.g.,][]{c66}
\begin{equation}
 \frac{d{\bf s}}{dt} = -\alpha \left({\bf c}\cdot{\bf s}\right)
  \left({\bf c} \times {\bf s}\right) , \label{sdyn1}
\end{equation}
where ${\bf c}$ denotes a unit vector along the orbital angular momentum, and
\begin{equation}
 \alpha = \frac{3}{2} \frac{GM_\odot}{\omega \,b^3}
  \frac{J_2+q}{\lambda + l}  \label{precc}
\end{equation}
is the precession constant of the planet. Here, $G$ is the gravitational
constant, $M_\odot$ is the mass of the Sun, $b=a\sqrt{1-e^2}$, where
$a$ and $e$ are the orbital semimajor axis and eccentricity,
$J_2$ is the quadrupole coefficient of the planetary gravity
field, and $\lambda=C/MR^2$ is the planetary moment of inertia
$C$ normalized by a standard factor $MR^2$, where $M$ is planet's
mass and $R$ its reference radius (to be also used in the definition
of $J_2$). The term $q=\case12 \sum_j (m_j/M) (a_j/R)^2$ is an
effective, long-term contribution to the
quadrupole coefficient due to the massive, close-in regular satellites
with masses $m_j$ and planetocentric distances $a_j$, and
$l=\sum_j (m_j a_j^2 n_j)/(M R^2\omega)$ is the angular momentum
content of the satellite orbits ($n_j$ denotes their planetocentric
mean motion) normalized to the characteristic value of the planetary
rotational angular momentum.

For both Jupiter and Saturn, $q$ is slightly dominant over $J_2$ in the numerator
of the last term in Eq.~(\ref{precc}), while $l$ is negligible in comparison
to $\lambda$ in the denominator. Spacecraft observations have been used to
accurately determine the values of $J_2$, $q$ and $l$. On the other hand,
$\lambda$ cannot be derived in a straightforward manner from observations,
because it depends on the structure of planetary interior. Using models of
planetary interior, \citet{hetal11} determined that Jupiter's $\lambda_{\rm J}$
is somewhere in the range between $0.2513$ and $0.2529$ (here rescaled for
the equatorial radius $R=71,492$~km of the planet). This would imply
Jupiter's precession constant $\alpha_{\rm J}$ to be in the range between
$2.754$~arcsec yr$^{-1}$ and $2.772$~arcsec yr$^{-1}$. These values
are smaller than the ones considered in \citet{wc06}, if their proposed
small angular distance between Jupiter's pole and Cassini state C$_2$ is
correct.

Similarly, \citet{hetal09} determined Saturn's precession constant
$\alpha_{\rm S}$ to be in the range between $0.8443$~arcsec yr$^{-1}$ and
$0.8447$~arcsec yr$^{-1}$. Again, these values differ from those inferred
by \citet{wh04} and \citet{hw04}, if a resonant confinement of the
Saturn's spin axis in their proposed scenario is true. This difference
has been noted and discussed in \citet{betal09}. If Helled's values are
correct, Saturn's spin axis cannot be presently locked in the resonance
with $s_8$. However, it seems possible that of $\alpha_{\rm S}$ derived in
\citet{hetal09} may have somewhat larger uncertainty than reflected by
the formal range of the inferred $\lambda$ values. Also, the interpretation
is complicated by the past orbital evolution of planetary satellites which
may have also contributed to changes of $\alpha_{\rm S}$ (and $\alpha_{\rm J}$).
For these reason, and in the spirit of previous studies
\cite[e.g.,][]{wh04,wc06,betal09}, here we consider a wider range of
the precession constant values $\alpha$ for both Jupiter and Saturn.

Assuming $\alpha$ constant for a moment, a difficult element preventing
a simple solution of Eq.~(\ref{sdyn1}) is the time evolution of
${\bf c}$. This is because mutual planetary interactions make their
orbits precess in space on a characteristic timescale of tens of
thousands of years and longer. In addition, during the early phase of
planetary evolution, the precession rates of planetary orbits may have been
faster due to the gravitational torques from a massive population of planetesimals
in the trans-Neptunian disk. In the Keplerian parametrization of orbits, the
unit vector ${\bf c}$ depends on the inclination $I$ and longitude of node
$\Omega$, such that ${\bf c}^T =(\sin I\sin\Omega,-\sin I\cos\Omega,\cos I)$.
Traditionally, the difficulties with the time dependence in Eq.~(\ref{sdyn1})
are resolved using a transformation to a reference frame fixed on an orbit,
where ${\bf c}^T =(0,0,1)$.

The transformation from the inertial to orbit coordinate frames is achieved by
applying a 3-1-3 rotation sequence with the Eulerian angles $(\Omega,I,-\Omega)$.
This transforms Eq.~(\ref{sdyn1}) to the following form
\begin{equation}
 \frac{d{\bf s}}{dt} = -\left[\alpha \left({\bf c}\cdot{\bf s}\right)
  {\bf c}+{\bf h}\right] \times {\bf s}, \label{sdyn2}
\end{equation}
where now the planetary spin vector ${\bf s}$ is expressed with
respect to the orbit frame, and ${\bf c}$ is now a unit vector
along the z-axis. In effect, the time dependence has been moved to the vector
quantity ${\bf h}^T =(\mathcal{A}, \mathcal{B},-2\mathcal{C})$ with
\begin{mathletters}
\begin{eqnarray}
 {\mathcal A} & \!\!\! = \!\!\!& \cos \Omega\,{\dot I} -\sin I\sin\Omega \,
  {\dot \Omega} , \label{abc1a} \\
 {\mathcal B} &\!\!\! =\!\!\! & \sin \Omega\,{\dot I} +\sin I\cos\Omega \,
  {\dot \Omega} , \label{abc1b} \\
 {\mathcal C} &\!\!\! =\!\!\! & \sin^2 I/2\,{\dot \Omega} , \label{abc1c}
\end{eqnarray}
\end{mathletters}
and the over-dots mean time derivative.

A further development consists in introducing complex and non-singular
parameter $\zeta = \sin(I/2)\,\exp(\imath\Omega)$ ($\imath=\sqrt{-1}$ is
the complex unit) that replaces $I$ and $\Omega$. First
order perturbation theory for quasi-circular and near-coplanar orbits
indicates $\zeta$ for each planet can be expressed using a finite number
of the Fourier terms with the $s_i$ frequencies uniquely dependent on
the orbital semimajor axes and masses of the planets, and amplitudes
set by the initial conditions \cite[e.g.,][]{bc61}. In the models
from non-linear theories or numerical integrations,  $\zeta$ can still be
represented by the Fourier expansion $\zeta=\sum A_i \exp[\imath(s_it+\phi_i)]$
\cite[e.g.,][]{aetal86,l88}, with the linear terms having typically the largest
amplitudes $A_i$.

As discussed in Section~1, terms with present frequencies $s_7\simeq -2.985$~arcsec
yr$^{-1}$ and $s_8\simeq -0.692$~arcsec yr$^{-1}$ are of a particular importance
in this work. The $s_7$-term is the largest in Uranus' $\zeta$ representation,
and the $s_8$-term is the largest in Neptune's $\zeta$ representation,
and these terms also appear, though with smaller amplitudes, in the $\zeta$
variable of Jupiter and Saturn, because mutual planetary interactions
enforce all fundamental frequency terms to appear in all planetary orbits.
In terms of $\zeta$, Eqs.~(4a-c) become
\begin{mathletters}
\begin{eqnarray}
 \mathcal{A} + \imath\, \mathcal{B} &\!\!\! =\!\!\! & \frac{2}{\sqrt{1-\zeta {\bar \zeta}}}
  \left(\frac{d\zeta}{dt}-\imath\zeta\,\mathcal{C}\right), \label{abc2a} \\
 \mathcal{C} &\!\!\! =\!\!\! & \frac{1}{2\imath} \left({\bar \zeta}\,\frac{d\zeta}{dt} -
   \zeta\, \frac{d{\bar \zeta}}{dt}\right), \label{abc2b}
\end{eqnarray}
\end{mathletters}
where ${\bar \zeta}$ is complex conjugate to $\zeta$. For small inclinations,
relevant to this work, we therefore find that $\mathcal{A} + \imath\, \mathcal{B}\simeq
2\,(d\zeta/dt)$, and $\mathcal{C}\simeq 0$.

Another important aspect is of the problem is that Eq.~(\ref{sdyn2})
derives from a Hamiltonian
\begin{equation}
 \mathcal{H}({\bf s};t) = \frac{\alpha}{2}\left({\bf c}\cdot
  {\bf s}\right)^2 + {\bf h}\cdot{\bf s} , \label{ham1}
\end{equation}
such that
\begin{equation}
 \frac{d{\bf s}}{dt} = -\nabla_{\bf s} \mathcal{H}\times {\bf s}.
  \label{sham2}
\end{equation}
This allowed \citet{betal05} to construct an efficient Lie-Poisson
integrator for a fast propagation of the secular evolution of
planetary spins. In Section~\ref{res}, we will use the leap-frog variant
of the Hamiltonian's LP2 splitting from \citet{betal05}. To propagate
${\bf s}$ through
a single integration step, \citet{betal05} method requires that the
orbital semimajor axis, eccentricity and ${\bf c}$ are provided at
times corresponding to the beginning and end of the step. These values can be
supplied from an analytic model of orbit evolution, or be directly obtained
from a previous numerical integrations of orbits where $a$, $e$ and
${\bf c}$ were recorded with a conveniently dense sampling.

\subsection{Parametrization using obliquity and precession angle}
The Hamiltonian formulation in Eq.~(\ref{ham1}) allows us to introduce
several important concepts of the Cassini dynamics. A long tradition in
astronomy is to represent ${\bf s}$ with obliquity $\varepsilon$
and precession angle $\psi$ such that ${\bf s}^T=(\sin\varepsilon
\sin\psi, \sin\varepsilon\cos\psi,\cos\varepsilon)$. The benefit
of this parametrization is that the unit spin vector is expressed
using only two variables. The drawback is that resulting equations
are singular when $\varepsilon=0$.

The conjugated  momentum to $\psi$ is $X=\cos\varepsilon$. The Hamiltonian is
then \cite[e.g.,][]{lr93}
\begin{eqnarray}
 \mathcal{H}(X,\psi;t) &=& \frac{\alpha}{2}X^2 - 2\mathcal{C} X
  \label{ham2} \\
  & & +\sqrt{1-X^2}\left(\mathcal{A}\sin\psi+ \mathcal{B}\cos\psi\right)
  . \nonumber
\end{eqnarray}
For the low-inclination orbits, $\mathcal{C}$ is negligible, while $\mathcal{A}$
and $\mathcal{B}$ are expanded in the Fourier series with the same frequency terms
as those appearing in $\zeta$ itself. A model of fundamental importance, introduced by
G.~Colombo \citep{c66,hm87,wh04}, is obtained when only one Fourier term
in $\mathcal{A}$ and $\mathcal{B}$ is considered. This Colombo model obviously
serves only as an approximation of the complete spin axis evolution, since all
other Fourier terms in $\mathcal{A}$ and $\mathcal{B}$ act as a perturbation.
Nevertheless, the Colombo model allows us to introduce several important concepts
that are the basis of the discussion in Section~\ref{res}.

In the Colombo model, the orbital inclination is fixed and the node
precesses with a constant frequency. Put in a compact way, $\zeta=A\exp[\imath(st+\phi)]$
is the single Fourier term, and $A=\sin I/2$ and $\Omega=st+\phi$. Transformation
to new canonical variables $X'=-X$ and $\varphi=-(\psi+\Omega)$, and scaling with
the nodal frequency $s$, results in a time-independent Hamiltonian
\begin{eqnarray}
 \mathcal{H}(X',\varphi) &=& \kappa X'^2 - \cos I X'
  \label{ham3} \\
 & & +\sin I \sqrt{1-X'^2}\cos\varphi ,  \nonumber
\end{eqnarray}
where $\kappa=\alpha/(2s)$ is a dimensionless parameter. Note that the
orbit-plane angle $\varphi$ is measured from a reference direction that is
$90^\circ$ ahead of the ascending node. The general structure of the phase
flow of solutions $\mathcal{H}(X',\varphi)=C$, with $C$ constant, derives
from the location of the stationary points. Depending on the parameter
values $(\kappa,I)$, there are either two ($|\kappa|<\kappa_\star$) or
four ($|\kappa|>\kappa_\star$) such stationary solutions (called the
Cassini states). The critical value of $\kappa$ reads \cite[e.g.,][]{hm87}
\begin{equation}
 \kappa_\star(I)=\frac{1}{2} \left(\sin^{2/3}I +\cos^{2/3}I\right)^{3/2} .
   \label{kk}
\end{equation}
Therefore, for small $I$, $\kappa_\star\simeq \case12$. The stationary
solutions are located at $\varphi=0^\circ$ or $\varphi=180^\circ$ meridians in the
orbital frame, and have obliquity values given by a transcendental equation
\begin{equation}
 \kappa \sin 2\varepsilon = -\sin\left(\varepsilon\mp I\right) ,
  \label{cs}
\end{equation}
with the upper sign for $\varphi=0^\circ$, and the lower sign for
$\varphi=180^\circ$.

In the present work, we are mainly interested in the Cassini state C$_2$ located
at $\varphi=0^\circ$. For Jupiter, the C$_2$ state related to frequency $s=s_7$
is subcritical since $|\kappa|< \case12$ for all estimates of the
Jupiter's precession constant found in the literature. Only two Cassini states
exist in this regime, and ${\bf s}$ must circulate about C$_2$. In the case of
Saturn, $|\kappa|> \kappa_\star\simeq \case12$ for $s=s_8$, four Cassini
states exist in this situation, and ${\bf s}$ was suggested to librate in
the resonant zone about C$_2$. The configuration of vectors in C$_2$ can be inferred
from Eq.~(\ref{cs}). If the inclination is significantly smaller than the obliquity
$\varepsilon$, we have that $\alpha\cos\varepsilon \simeq -s$. Since the term on
the left hand side of this relation is the precession frequency of the planet's
spin (see Eq.~\ref{sdyn1}), we find that the spin and orbit vectors will co-precess
with the same rate about the normal vector to the inertial frame. Small resonant
librations about C$_2$ would reveal themselves by small departures of the spin vector
from this ideal state. The maximum width $\Delta\varepsilon$ of the resonant zone
in obliquity at the $\varphi=0^\circ$ meridian can be obtained using an
analytic formula \cite[e.g.,][]{hm87}
\begin{equation}
 \sin\frac{\Delta\varepsilon}{2} = \frac{1}{|\kappa|} \sqrt{\frac{\sin 2I}{\sin
  2\varepsilon_4}} , \label{rw}
\end{equation}
where $\varepsilon_4$ is the obliquity of the unstable Cassini state C$_4$ (a solution
of Eq.~(\ref{cs}) at $\varphi=180^\circ$ having the intermediate value of the
obliquity).

\begin{figure*}[t]
\epsscale{1.0}
\plotone{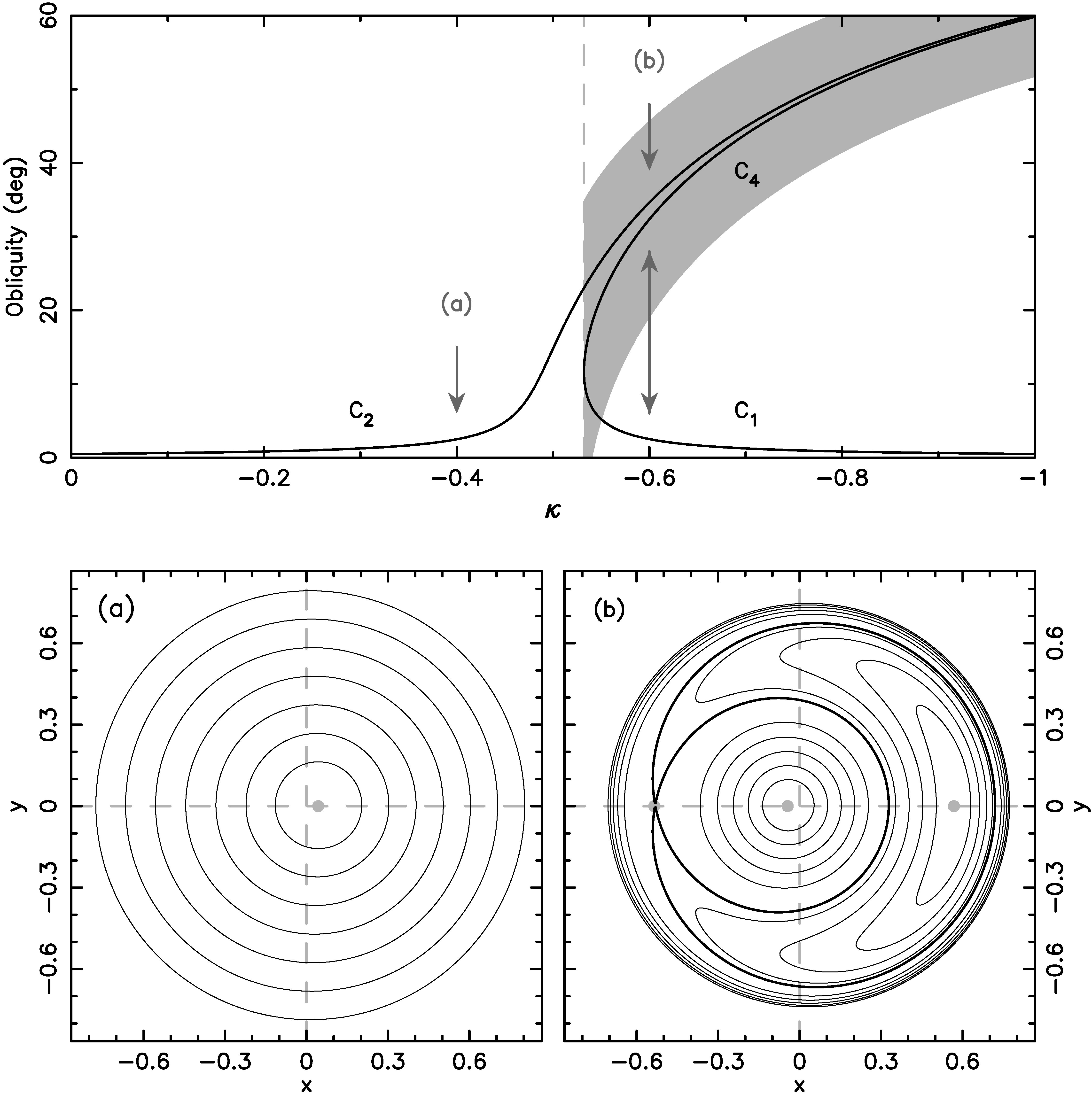}
\caption{Top panel: Parametric dependence of Cassini state C$_1$, C$_2$ and C$_4$
 obliquity (ordinate) on the frequency-ratio parameter $\kappa$ (abscissa)
 in the Colombo model (the C$_3$ equilibirium has obliquity larger than $90^\circ$
 and it is not relevant for our discussion). The orbital inclination $I$ has been
 set to $0.5^\circ$ value. The dashed line indicates the critical value
 $-\kappa_\star$ at which C$_1$ and C$_4$ bifurcate (Eq.~10). The gray area for
 $|\kappa|> \kappa_\star$ shows maximum width of the resonant zone around C$_2$.
 Bottom panels: Examples of phase portraits ${\cal H}(X',\varphi)=C$ for
 two values of $\kappa$: (a) $\kappa=-0.4$ on the left, and (b) $\kappa=-0.6$
 on the right. We use coordinates $x=\sin\varepsilon
 \cos\varphi$ and $y=\sin\varepsilon\sin\varphi$ with the origin at the north
 pole $\varepsilon=0$. The gray symbols show location of the Cassini
 equilibria and the curves are isolines ${\cal H}(X',\varphi)=C$ for
 suitably chosen $C$ values. The bold line in (b) is the separatrix of the
 resonant zone around C$_2$ stationary solution.
\label{fig1}}
\end{figure*}

Figure~\ref{fig1}, top panel, shows how the location of the Cassini states and the
resonance width $\Delta\varepsilon$ depend on $\kappa$, which is
the fundamental parameter that changes during planetary migration. For sake of
this example we assumed orbital inclination $I=0.5^\circ$ (note that the overall
structure remains similar for even smaller inclination values considered in the
next section, but would be only less apparent in the Figure). The C$_1$ and
C$_4$ stationary solution bifurcate when $\kappa=\kappa_\star$ at a non-zero critical
obliquity value $\varepsilon_\star={\rm atan}(\tan^{1/3} I)$ \cite[e.g.,][]{hm87,wh04}.
Note that $\Delta\varepsilon$ is significant in spite of a very small value of
the inclination, which manifests through its dependence on a square
root of $\sin 2I$ in (\ref{rw}). The bottom panels show examples of the
phase portraits ${\cal H}(X',\varphi)=C$ for both sub-critical $|\kappa|<\kappa_\star$
and super-critical $|\kappa|>\kappa_\star$ cases.

\section{Results} \label{res}
We now turn our attention to the evolution of Jupiter's and Saturn's obliquities
during planetary migration. We first discuss the orbital evolution of planets
in the instability/migration simulations of NM12 (Section~3.1). We then
parametrize the planetary migration before (stage~1) and after the instability
(stage~2), and use it to study the effects on Jupiter's and Saturn's obliquities.
The two stages are considered separately in Sections~3.2 and 3.3.

\subsection{The orbital evolution of giant planets}
NM12 reported the results of nearly $10^4$ numerical integrations of planetary
instability, starting from hundreds of different initial configurations of planets
that were obtained from previous hydrodynamical and $N$-body calculations.
The initial configurations with the 3:2 Jupiter-Saturn mean motion resonance
were given special attention, because Jupiter and Saturn, radially migrating in
the gas disk before its dispersal, should have become trapped into their mutual
3:2 resonance \cite[e.g.,][]{ms01,mc07,pn08}. They considered cases with four,
five and six
initial planets, where the additional planets were placed onto resonant orbits
between Saturn and the inner ice giant, or beyond the orbit of the outer ice giant.
The integrations included the effects of the transplanetary planetesimal disk.
NM12 experimented with different disk masses, density profiles, instability
triggers, etc., in an attempt to find solutions that satisfy several constraints,
such as the orbital configuration of the outer planets, survival of the
terrestrial planet, and the distribution of orbits in the asteroid belt.

NM12 found the dynamical evolution in the four planet case was typically too
violent if Jupiter and Saturn start in the 3:2 resonance, leading to ejection of
at least one ice giant from the Solar System. Planet ejection can be avoided if
the mass of the transplanetary disk of planetesimals was large ($M_{\rm disk}
\gtrsim 50$ $M_{\rm Earth}$, where $M_{\rm Earth}$ is the Earth mass), but
such a massive disk would lead to excessive dynamical damping (e.g., the
outer planet orbits become more circular then they are in reality) and to smooth
migration that violates constraints from the survival of the terrestrial planets,
and the asteroid belt. Better results were obtained when the Solar System
was assumed to have five giant planets initially and one ice giant, with mass
comparable to that of Uranus or Neptune, was ejected into interstellar space
by Jupiter \citep{n11,betal12}. The best results were obtained when the
ejected planet was placed into the external 3:2 or 4:3 resonances with Saturn
and $M_{\rm disk}\simeq 20$~$M_{\rm Earth}$. The range of possible outcomes was
rather broad in this case, indicating that the present Solar System
is neither a typical nor expected result for a given initial state.

\begin{figure*}[t]
\epsscale{1.0}
\plotone{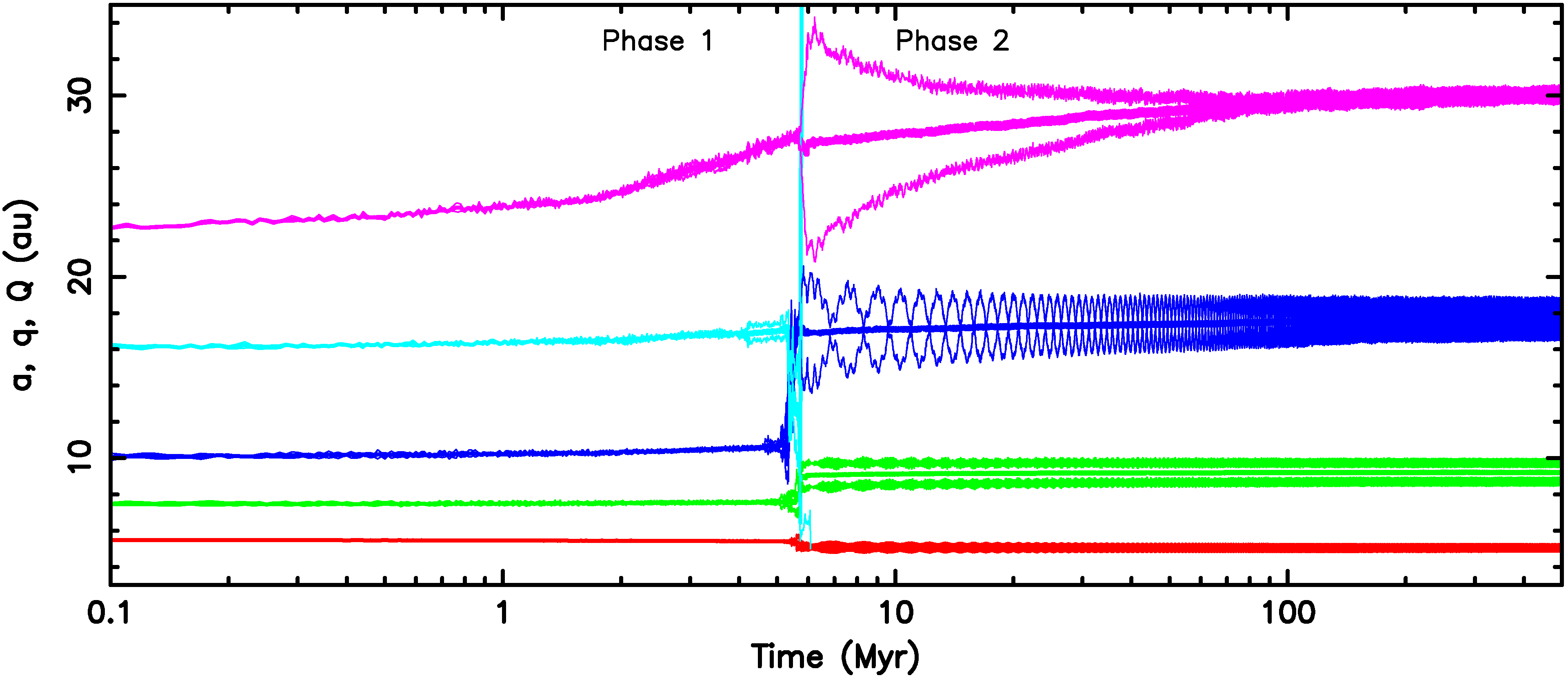}
\caption{An example of planetary migration and instability from \citet{nm12}.
 The plot shows the evolution of the semimajor axis (bold line), and the
 perihelion and aphelion distances (thin lines) of the giant planets. The
 initial orbits of Jupiter, Saturn and the inner ice giant were placed in
 the 3:2 resonant chain. The semimajor axes of the two outer ice giants were
 set to be $16$~au and $22$~au. The trans-Neptunian disk of planetesimals
 (not shown here) was resolved by 10,000 equal-mass particles. The disk
 originally extended from $23.5$~au to $29$~au and had the total mass of
 $20$ $M_{\rm Earth}$. The instability happened at $t\simeq 5.6$~Myr after
 the start of the simulation. The third ice giant was subsequently ejected
 from the Solar system. Note that the migration rate before the instability
 (phase~1) is significantly larger than the migration rate after the
 instability (phase~2).
\label{fig2}}
\end{figure*}

The most relevant feature of the NM12 models for this work is that the
planetary migration happens in two stages (see Fig.~\ref{fig2}). During the
first stage, that is before the instability happens, Neptune migrates into
the outer disk at $\simeq 20-30$~au. The migration is relatively fast during
this stage, because the outer disk still has a relatively large mass. We
analyzed several simulations from NM12 and found that Neptune's migration
can be approximately described by an exponential with the e-folding
timescale $\tau\simeq 10$~Myr for $M_{\rm disk}=20$ $M_{\rm Earth}$ and
$\tau\simeq 20$ Myr for $M_{\rm disk}=15$ $M_{\rm Earth}$. The instability typically
happens in the NM12 models when Neptune reaches $\simeq 28$~au. The main
characteristic of the instability is that planetary encounters occur, mainly
between the extra ice giant and all other planets. The instability typically
lasts $\sim 10^5$ years and terminates when the extra ice giant is ejected from
the solar system by Jupiter. The second stage of migration starts after that.
The migration of Neptune is much slower during this period, because the outer
disk is now very much depleted. From simulations in NM12 we find that
$\tau\simeq 30$~Myr for $M_{\rm disk}=20$ $M_{\rm Earth}$ and $\tau\simeq 50$~Myr for
$M_{\rm disk}=15$ $M_{\rm Earth}$. Moreover, rather then being precisely
exponential, the migration slows down relative to an exponential with fixed
$\tau$, such as, effectively, the very late stages have larger $\tau$ values
($\tau\sim 100$~Myr) than the ones immediately following the instability.
Uranus accompanies the migration of Neptune on timescales similar
to those mentioned above.

The frequencies $s_7$ and $s_8$, which are the most relevant for this work,
are initially somewhat higher than the precession constants of Jupiter and
Saturn, mainly because of the torques from the outer disk. The extra term from
the third ice giant initially located at $\simeq 10$~au has much faster
frequency then the precession constants and does not interfere with the
obliquity of Jupiter and Saturn during the subsequent evolution. The $s_7$
and $s_8$ frequencies slowly decrease during both stages. Their e-folding
timescales may slightly differ from the migration e-folding timescales
mentioned above, due to the non-linearity of the dependence of the secular
frequencies on the semimajor axis of planets. Our tests show that they are
about $(90-95)$\% of the e-folding timescales of planetary semimajor axes.

From analyzing the behavior of frequencies in different simulations we found
that $s_8$ should cross the value of Jupiter's precession constant during the
first migration stage, that is {\it before} the instability. The main
characteristic of this crossing is that the planetary orbits are very nearly
coplanar during this stage. The amplitude $I$ in Eq.~(\ref{ham3}) should thus
be very small. It is not known exactly, however, how small. In Section~3.2,
we consider amplitudes down to $0.005^\circ$ (about $10$ times smaller than the
present value of $I_{58}$, where $I_{58}$ is the amplitude of the $8$th frequency
term in Jupiter's orbit), and show that the effects on Jupiter's obliquity
are negligible if the amplitudes were even lower. The second characteristic
of the first migration stage is that the evolution of $s_8$ happens on a
characteristic timescale of $\simeq 10-20$~Myr. Since the total change of $s_8$
during this interval is several arcsec~yr$^{-1}$, and the first stage typically lasts
$\sim 10-20$~Myr, the average rate of change is very roughly, as an order of
magnitude estimate, ${\rm d}s_8/{\rm d}t\sim 0.1$ arcsec~yr$^{-1}$~Myr$^{-1}$.
The actual value of ${\rm d}s_8/{\rm d}t$ during crossing depends on several
unknowns, including when exactly the crossing happens during the first stage.
Also, the changes of $s_8$ could have been slower if the first stage lasted
longer than in the NM12 simulations, as required if the instability
occurred at the time of the Late Heavy Bombardment \cite[e.g.,][]{getal05}.
In Section~3.2, we will consider values in the range $0.005<{\rm d}s_8/{\rm d}t<
0.05$ arcsec~yr$^{-1}$~Myr$^{-1}$, and show that the obliquity of Jupiter
cannot be pumped up to its current value if ${\rm d}s_8/{\rm d}t>0.05$
arcsec~yr$^{-1}$~Myr$^{-1}$ (assuming that $I_{58}\lesssim 0.05^\circ$).

Interesting effects on obliquities should happen during the second migration
stage. First, the $s_8$ frequency reaches the value of the precession constant
of Saturn. There are several differences with respect to the $s_8$ crossing
of Jupiter's precession constant during the first stage (discussed above).
The orbital inclinations of planets were presumably excited to their current
values during the instability. Therefore, the amplitude $I_{68}$ should be
comparable to its current value, $I_{68} \simeq 0.064^\circ$, during the
second stage. We see this happening in the NM12 simulations. First, there
is a brief period during the instability, when the inclinations of all
planets are excited by encounters with the ejected ice giant. The inclination
of Neptune is modest, at most $\simeq2^\circ$, and is rather quickly damped
by the planetesimal disk. Also, the invariant plane of the solar system
changes by $\simeq 0.5^\circ$ when the third ice giant is ejected during
the instability. The final inclinations are of this order. The current
amplitudes are $I_{\rm 58} \simeq 0.066^\circ$ ($s_8$ term in Jupiter's orbit)
and $I_{\rm 68}\simeq 0.064^\circ$ ($s_8$ term in Saturn's orbit)
\cite[see e.g.,][]{l88}.

Another difference with respect the first stage is that the evolution of
$s_8$ is much slower during the second stage. If, as indicated by the NM12
integration, $s_8$ changes by $\sim 1$~arcsec~yr$^{-1}$ in $100$~Myr, then
the average rate of change is very roughly ${\rm d}s_8/{\rm d}t\sim0.01$
arcsec~yr$^{-1}$~Myr$^{-1}$. The actual rate of change can be considerably
lower than this during the very late times, when the effective $\tau$ was
lower than during the initial stages. Finally, during the very last gasp of
migration, the $s_7$ frequency should have approached the precession constant
of Jupiter. We study this case in an adiabatic approximation when the rate
of change of $s_7$ is much slower than any other relevant timescale. We find
that the present obliquity of Jupiter can be excited by the interaction with
the $s_7$ term only if the precession constant of Jupiter is somewhat larger
than inferred by \cite{hetal11}, in accord with the results of \cite{wc06}.

\subsection{The effects on Jupiter's obliquity during stage~1}
Since $s_8$ remains larger than $\alpha_{\rm S}$ during the first stage, we
do not expect any important effects on Saturn's obliquity during this stage.
If Saturn's obliquity was low initially, it should have remained low in
all times before the instability. We therefore focus on the case of Jupiter
in this section. From the analysis of the NM12 numerical simulation in
Section~3.1, we infer that the $s_8$ frequency crossed $\alpha_{\rm J}$ during
the first stage. The values of ${\rm d}s_8/{\rm d}t$ and $I_{58}$ during
crossing are not known exactly from the NM12 simulations, because they
depend on details of the initial conditions. We therefore consider a range
of values and determine how Jupiter's obliquity excitation depends on them.
The results can be used to constrain future simulations of the planetary
instability/migration.

We consider Colombo's model with only one Fourier term in $\zeta$ of Jupiter,
namely that of the $s_8$ frequency.%
\footnote{We found that adding higher frequency terms, such as $s_6$ 
 and/or $s_7$, into our simulation does not change results.}
The inclination term $I_{58}$ in Jupiter's orbit was treated as a free
parameter. The range of values was set to be between zero and 
roughly the current
value of $\simeq 0.066^\circ$. As we discussed in Section~3.1, it is
reasonable to assume that $I_{58}$ was smaller than the current value,
because the orbital inclination of planets should have been low in times
before the instability. The value of $\alpha_{\rm J}$ was obtained by rescaling
the present value to the the semimajor axis of Jupiter before the instability
($a_{\rm J} \simeq 5.45$~au). To do so we used Eq.~(\ref{precc}) and
assumed $\alpha_{\rm J} \propto a_{\rm J}^{-3}$. No additional modeling of
possible past changes of $\alpha_{\rm J}$, for instance due to satellite system
evolution or planetary contraction, was implemented. The $s_8$ frequency
was slowly decreased from a value larger than $\alpha_{\rm J}$ to a value
smaller than $\alpha_{\rm J}$.

Motivated by the numerical simulations of the instability discussed in
Section~3.1, we assumed the initial $s_8$ value of $-4$~arcsec yr$^{-1}$ and
let it decrease to $-1.2$~arcsec yr$^{-1}$ by the end of each test. The rate
of change, ${\rm d} s_8/{\rm d}t$, was treated as a free parameter. The
integrations were carried for several tens of Myr for the highest assumed
rates and up to several hundreds of Myr for the lowest rates. We recorded
Jupiter's obliquity during the last $5$~Myr of each run, and computed the
mean value $\varepsilon_{\rm fin}$. Jupiter's initial spin axis was oriented
toward the pole of the Laplacian plane. [The value of $\varepsilon_{\rm fin}$
reported in Fig.~\ref{fig3} was averaged over all possible phases of the
initial spin axis on the $\mathcal{H}(X',\varphi)=C$ level curve, with $C$
defined by ${\bf s}$ oriented toward the pole of the Laplacian plane]

Figure~\ref{fig3} shows the results. For most parameter combinations shown here
the $s_8$ resonance swept over $\alpha_{\rm J}$ without having the ability
to capture Jupiter's spin vector in the resonance. This happened because
${\rm d}s_8/{\rm d}t$ was relatively large and $I_{58}$ was relatively small,
thus implying that the $s_8$ frequency crossed the resonant zone in a time
interval that was shorter than the libration period. Captures occurred only in
extreme cases (largest $I_{58}$ and smallest $ds_8/dt$). These cases ended
up generating very large obliquity values of Jupiter and are clearly implausible.
The plausible values of ${\rm d}s_8/{\rm d}t$ and $I_{58}$ correspond to the cases
where Jupiter's obliquity was not excited at all, thus assuming that Jupiter
obtained its present obliquity later, or was excited by up to $\simeq 3^\circ$. To
obtain $\varepsilon_{\rm J}\simeq 3^\circ$, ${\rm d}s_8/{\rm d}t$ and $I_{58}$ would
need to have values along the bold line labeled~3 in Fig.~\ref{fig3}, which extends
diagonally in ${\rm d}s_8/{\rm d}t$ and $I_{58}$ space. An example of a case,
where the obliquity of Jupiter was excited to near $\simeq 3^\circ$ value, is
shown in Fig.~\ref{fig4}.

\subsection{Behavior of obliquities during stage~2}
We now turn our attention to the second stage, when the migration slowed
down and the obliquities of Jupiter and Saturn should have suffered
additional perturbations. At the beginning of stage, that is just after the
time of the instability, the $s_8$ frequency is already lower than
$\alpha_{\rm J}$, but still higher than $\alpha_{\rm S}$, while the $s_7$
frequency is higher than $\alpha_{\rm J}$. Since the $s_7$ and $s_8$
frequencies are slowly decreasing during the second stage, a possibility
arises that Jupiter's obliquity was (slightly) excited when $s_7$ approached
$\alpha_{\rm J}$ \cite[e.g.,][]{wc06}, and that Saturn's obliquity was strongly
excited by capture into the spin-orbit resonance with $s_8$
\cite[e.g.,][]{wh04,hw04,betal09}.

\subsubsection{Jupiter}
\cite{wc06} suggested a possibility that Jupiter's present obliquity may
be explained by the proximity of $\alpha_{\rm J}$ to the current value of
the $s_7$ frequency. They showed that, if $\kappa=\alpha_{\rm J}/(2s_7)$ is
sufficiently close to the critical value from Eq.~(\ref{kk}), namely $\simeq
\case12$ for small inclinations, the obliquity of the Cassini state C$_2$ may
be significant. Thus, as $s_7$ adiabatically approached to $\alpha_{\rm J}$,
Jupiter's obliquity may have been excited along. As a supportive argument
for this scenario they pointed out that $\varphi_{\rm J} \simeq 0^\circ$, where
the Cassini state 2 is located.

To test this possibility we run a suite of simulations, assuming an
exponential convergence to the current value $s_7\simeq -2.985$~arcsec
yr$^{-1}$. Specifically, we set $s_7(t)=s_7(0)+[s_7-s_7(0)]\exp(-t/\tau)$,
where $\tau$ and $s_7(0)$ are parameters. The initial value $s_7(0)$ at the
beginning of the second stage was obtained from the numerical simulation
discussed in Section~3.1. Here we chose to use $s_7(0)=-3.5$~arcsec yr$^{-1}$,
however we verified that the results are insensitive to this choice.
The e-folding timescale $\tau$ depends on how slow or fast planets migrate.
Given that the planetary migration is slow during the second stage, we chose
$\tau=100-200$~Myr. This assures that the approach of $s_7(t)$ to $\alpha_{\rm J}$
is adiabatical. The amplitude $I_{57}$ is assumed to be constant and equal to
its current value ($I_{57} \simeq 0.055^\circ$).

Two additional parameters need to be specified: (i) Jupiter's initial spin
state, and (ii) $\alpha_{\rm J}$. As for (i), the results discussed in
Section~3.2 indicate that Jupiter's obliquity may have remained near zero
during the first stage, if $I_{58}$ was too small and/or ${\rm d}s_8/{\rm d}t$
was too fast, or could have been potentially excited to $\simeq 3^\circ$, if
$I_{58}$ and ${\rm d}s_8/{\rm d}t$ combined in the right way. Therefore, here we
treat the obliquity of Jupiter at the beginning of stage~2 as a free parameter.
As for (ii), as we discussed in Section~1, the present value of $\alpha_{\rm J}$
somewhat uncertain. We therefore performed various simulations, where
$\alpha_{\rm J}$ takes on a number of different values between $2.75$
arcsec~yr$^{-1}$ and 2.95 arcsec~yr$^{-1}$. A similar approach has also been
adopted by \cite{wc06}.

Figure~\ref{fig5} reports the results. The top panel shows how the obliquity
$\varepsilon_{\rm C2}$ of the Cassini state C$_2$ depends on the assumed value
of $\alpha_{\rm J}$. This is calculated from Eq.~(\ref{cs}). The trend is that
$\varepsilon_{\rm C2}$ increases with $\alpha_{\rm J}$, because the larger values
of $\alpha_{\rm J}$ correspond to a situation where the system is closer to the
exact resonance with $s_7$. If $\alpha_{\rm J}<2.8$~arcsec yr$^{-1}$, as inferred
from models in \cite{hetal11}, $\varepsilon_{\rm C2}$ is too small to
significantly contribute to $\varepsilon_{\rm J}$. This case would imply
that Jupiter's present obliquity had to be acquired during the earlier stages
and is possibly related to the non-adiabatic $s_8$ resonance crossing discussed
in Section~3.2. If, on the other hand, $\alpha_{\rm J}\simeq 2.92-2.94$~arcsec
yr$^{-1}$, Jupiter would owe its present obliquity to the proximity between
$\alpha_{\rm J}$ and $s_7$. This would imply that the obliquity excitation
during the first stage of planetary migration must have been minimal.
Figures~\ref{fig3} and \ref{fig5} express the joint constraint on the
planetary migration also for the intermediate cases, where the present
obliquity of Jupiter arose as a combination of both effects discussed here.

Figure~\ref{fig6} illustrates the two limiting cases discussed above. In
panel (a), we assumed that the parameters during the first migration stage
were such that the obliquity of Jupiter was excited to its current value
during the $s_8$ crossing (such as shown in Fig.~\ref{fig4}). Also, we
set $\alpha_{\rm J}=2.77$~arcsec yr$^{-1}$, corresponding to the best theoretical
value of \cite{hetal11}. The Cassini state C$_2$ corresponding to the $s_7$ term
is only slightly displaced from the center of the plot, and does not
significantly contribute to the present obliquity value. In panel (b) of
Fig.~\ref{fig6}, we set $\alpha_{\rm J}=2.93$~arcsec yr$^{-1}$. This value implies
that $\varepsilon_{\rm C2}\simeq 2.6^\circ$. The present obliquity of Jupiter would
then be in large part due to the ``forced'' term arising from the proximity
of $s_7$. The initial excitation of Jupiter's obliquity during the first
stage would have to be minor in this case.

\subsubsection{Saturn}
In the case of Saturn, all action is expected to take place during the second
stage of planetary migration. Insights gleaned from the numerical simulations
discussed in Section~3.1 show that the $s_8$ frequency should have very slowly
approached $\alpha_{\rm S}$, thus providing a conceptual basis for capture of
Saturn's spin vector in a resonance with $s_8$ \cite[e.g.,][]{wh04,hw04,betal09}.
To study this possibility, we assume that the $I_{68}$ amplitude was excited
to its current value during the instability, and remained nearly constant during
the second stage of planetary migration. This choice is motivated by the NM12
simulations, where the inclination of Neptune is never too large. Note that
\cite{betal09} investigated the opposite case where Neptune's inclination was
substantially excited during the instability and remained high when the resonance
with $s_8$ occurred. This type of strong inclination excitation does not happen
in the NM12 models.

Here we assume that the planetary migration was very slow during the second
stage and parametrize $s_8(t)$ as $s_8(t)=s_8(0)+[s_8-s_8(0)]\exp(-t/\tau)$ with
$\tau\geq 80$~Myr. The initial frequency value at the beginning of stage~2, $s_8(0)$,
is estimated from the NM12 simulations. We find that $s_8(0)\simeq -1.3$~arcsec
yr$^{-1}$, and use this value to set up the evolution of $s_8(t)$. We also assume
a range of $\alpha_{\rm S}$ values. This has the following significance. As already
pointed out by \citet{betal09}, the best-modeling values of $\alpha_{\rm S}$ from 
\cite{hetal09} are not compatible with a resonant location of Saturn's spin axis. This
is because the Cassini state C$_2$ would be moved to a significantly larger
obliquity value ($\geq 34^\circ$). So these values of $\alpha_{\rm S}$ would imply
that Saturn's spin circulates about the Cassini state C$_1$. On the other hand,
the significant obliquity of Saturn requires an increase when the $s_8$ value
was crossing $\alpha_{\rm S}$ value, as schematically shown in the left panel
of Fig.~\ref{fig4} for Jupiter's obliquity during the phase~1. \citet{betal09}
tested this scenario using numerical simulations and found it extremely unlikely:
initial data of an insignificant measure have led this way to the current
spin state of Saturn. Indeed, here we recover the same result with a less
extensive set of numerical simulations.

Given the arguments discussed above we therefore tend to believe that the
precession constant of Saturn may be somewhat smaller than the one determined
by \cite{hetal09}. For instance, R.~A. Jacobson (personal communication) determined
the Saturn precession from the Saturn's ring observations. The mean precession
rate obtained by him is $0.725$ arcsec~yr$^{-1}$ (formal uncertainty of about
6\%). This value would indicate $\alpha_{\rm S}$ in the range between
$0.769$ arcsec~yr$^{-1}$ and $0.864$ arcsec~yr$^{-1}$. Both \cite{wh04} and
\cite{betal09} report other observational constraints of Saturn's pole
precession that have comparably large uncertainty. We therefore sampled a larger
interval of the $\alpha_{\rm S}$ values to make sure that all interesting
possibilities are accounted for.

Our numerical simulations thus spanned a grid of two parameters: (i) $\alpha_{\rm S}$
discussed above, and (ii) $\tau$, the e-folding timescale of the $s_8$
frequency that slowly changes due to residual migration of Neptune and depletion
of the outer disk. The amplitude
related to the $s_8$ term in the inclination vector $\zeta$ of Saturn is
kept constant, namely $I_{68}=0.064^\circ$. To keep number of tested free
parameters low, we assumed initial orientation of Saturn's spin axis ${\bf s}$
to be near the pole of the invariable plane. Specifically, we set its obliquity to
$0.1^\circ$ in the reference frame of the $s_8$-frequency Fourier term in
$\zeta$. To prevent fluke results, we sampled 36 values of longitude
$\varphi$ of the Saturn's pole in the same reference frame, incrementing it from
$0^\circ$ by $10^\circ$. Each of the simulations covered a $1$~Gyr timespan. 
We recorded Saturn's pole orientation during the last $150$~Myr time interval.
We specifically analysed if it passes close to the current location of Saturn's
pole, namely $\varepsilon_8\simeq 27.4^\circ$ and $\varphi_8\simeq -31.4^\circ$ 
in the $s_8$-frequency reference frame (see Table~2 in Ward \& Hamilton 2004).
A numerical run was considered successful, if the simulated
Saturn's pole passed through a box of $\pm 0.2^\circ$ in obliquity $\varepsilon$
and $\pm 3^\circ$ in longitude $\varphi$ around the planet's values
$(\varepsilon_8,\varphi_8)$ during the recorded $150$~Myr time interval. Note that
the libration period of Saturn's pole around Cassini state C$_2$, if captured
in the spin-orbit resonance, is $\sim (50-100)$~Myr, depending on the libration
amplitude. This set our requirement for the timespan over which we monitored
Saturn's pole position.

Figure~\ref{fig7} shows the results from this suite of runs. The shaded
region shows correlated $\alpha_{\rm S}$-$\tau$ pairs that provided successful
match to the Saturn's pole position. We note that no successful solutions
were obtained for $\alpha_{\rm S}>0.812$ arcsec~yr$^{-1}$ and all successful
solutions correspond to the capture in the resonance zone around the Cassini
state C$_2$. The absence of low-probability solutions in which Saturn's pole would
circulate about the Cassini state C$_1$ for larger $\alpha_{\rm S}$ may be
related to the limited number of simulations performed. No solutions
were also obtained for $\tau>215$~Myr. This is because for such long
e-folding timescales the resonant capture process would be strictly adiabatic
and the simulated spin would not meet the condition of at least $30^\circ$
libration amplitude \cite[see discussion in][]{wh04,hw04}. The area
occupied by successful solutions splits into two branches for $\alpha_{\rm S}
\simeq 0.78$ arcsec~yr$^{-1}$. This is because the obliquity of the Cassini state
C$_2$ is $\varepsilon_{\rm C2}\simeq 27.4^\circ$ for the critical value of
$\alpha_{\rm S}$, and the solutions have the minimum libration amplitude
of $\simeq 31^\circ$.

Figure~\ref{fig8} shows two evolution paths of Saturn's spin vector obtained
in two different simulations. As mentioned above, in both cases Saturn's spin
state was captured into the spin-orbit resonance $s_8$, and remained in the
resonance for the full length of the simulation. The final states of both
simulations are a good proxy for the Saturn's present spin state. These two cases
differ from each other principally because the path in panel (b) shows librations
with a larger amplitude than the path in panel (a). Note some of the solutions,
such as (b) here, may attain a significant librations amplitude. This is
because with the corresponding values of $\tau\simeq 100$~Myr the evolution
is not adiabatic and the librations amplitude is excited immediately after
capture. Therefore, the complicated evolution histories proposed in
\cite{hw04} may not be needed.

\section{Conclusions}
We studied the behavior of Jupiter's and Saturn's obliquities in models of
planetary instability and migration that were informed from NM12. Rather then
investigating a few specific cases directly from NM12, we considered the
general concept of a two-stage migration from NM12, and studied a broad range
of relevant parameter values. We found that, in general, the two stage migration
provides the right framework for an adequate excitation of Jupiter's and
Saturn's obliquities. Moreover, we found that certain conditions must be
satisfied during the first and second stages of migration, if the final
obliquity values are to match the present obliquities of these planets.

\begin{figure}[t]
\epsscale{0.8}
\plotone{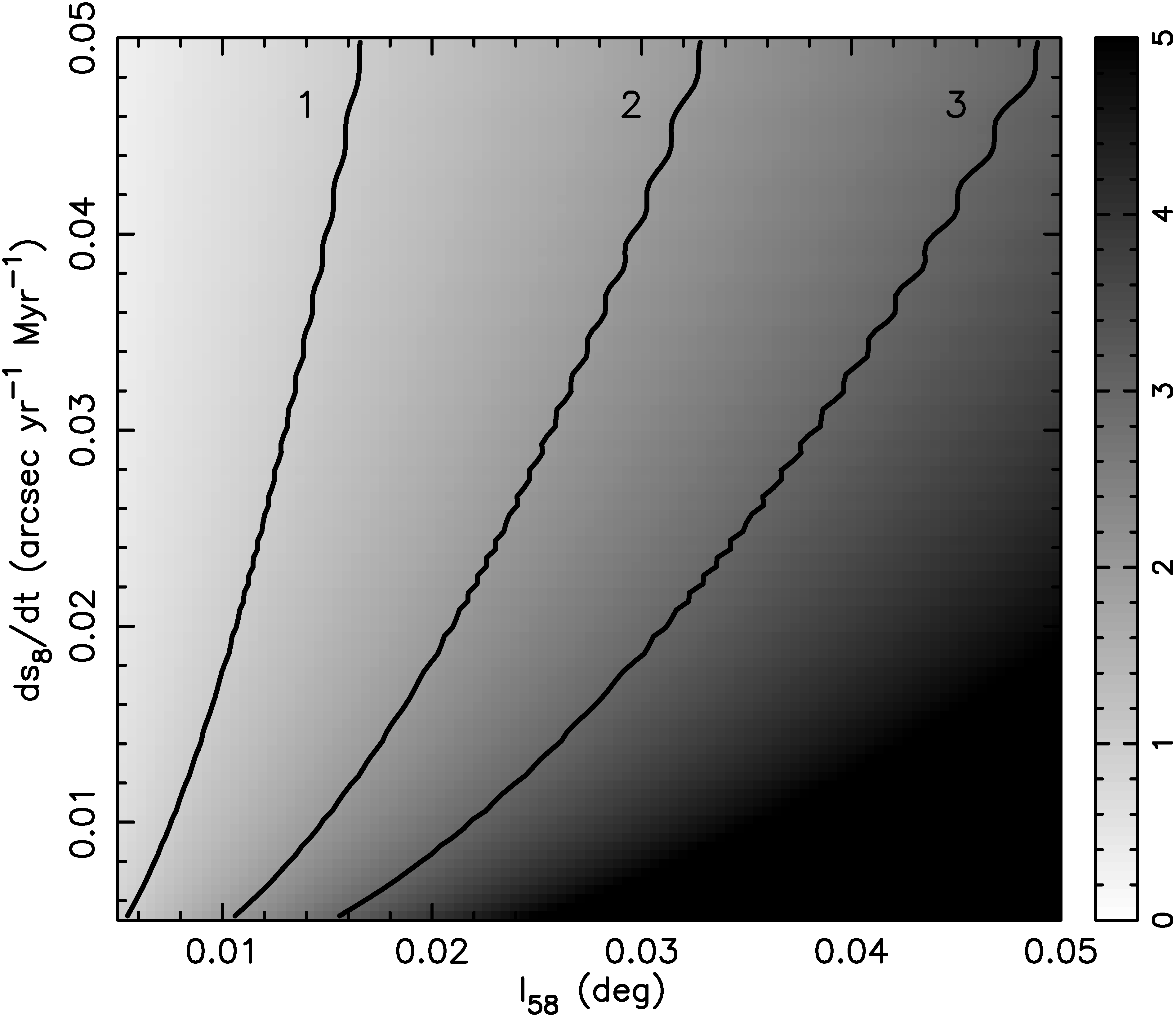}
\caption{The final obliquity $\varepsilon_{\rm fin}$ of Jupiter obtained in
 our integrations of
 crossing of the $s_8$ spin-orbit resonance. Jupiter's obliquity is
 shown as a function of two parameters: (i) the amplitude $I_{58}$ of the
 Fourier term in Jupiter's $\zeta$ variable (see Section~2), and
 (ii) the rate ${\rm d}s_8/{\rm d}t$ at the time when $s_8$ became
 equal to $\alpha_{\rm J}$. The gray shading indicates the final obliquity
 (the scale bar on the right shows the corresponding numerical value in
 degrees). The three bold isolines correspond to $\varepsilon_{\rm J}=1^\circ$,
 $2^\circ$ and $3^\circ$ (see the labels).
 \label{fig3}}
\end{figure}

\begin{figure*}[t]
\epsscale{1.0}
\plotone{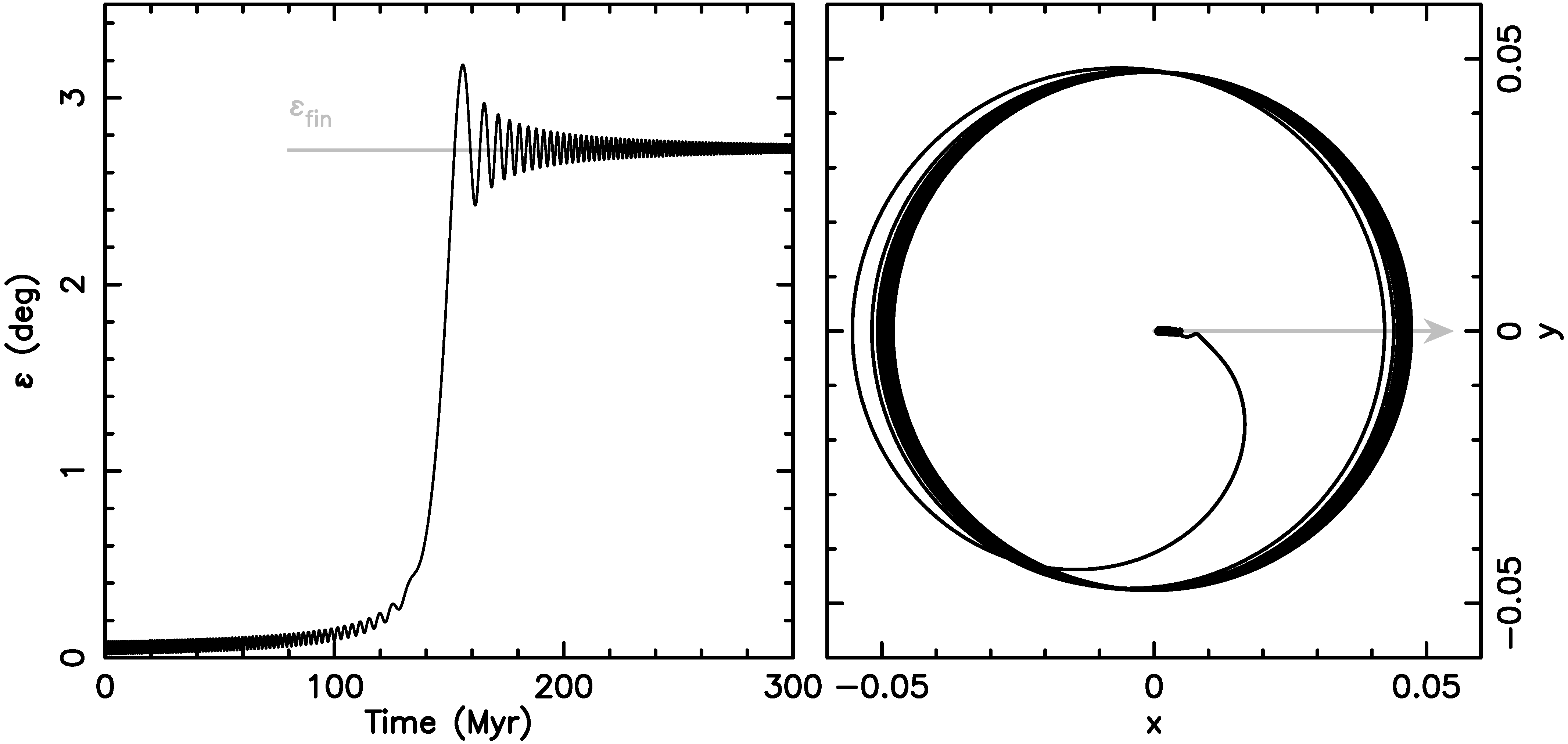}
\caption{An example demonstrating the effect of $s_8$ frequency sweeping
 over $\alpha_{\rm J}$. Here we set $I_{58}=0.02^\circ$ and ${\rm d}s_8/{\rm d}t
 =0.01$~arcsec yr$^{-1}$ Myr$^{-1}$. The left panel shows Jupiter's obliquity
 as a function of time. Obliquity $\varepsilon_{\rm J}$ increases to $\simeq
 2.7^\circ$ during the resonance crossing. The width of the Cassini resonance
 is small because $I_{58}$ is small, and the assumed rate ${\rm d}s_8/{\rm d}t$
 is too large in this case to lead to capture. The right panel shows the spin
 axis evolution projected onto the $(x,y)$ plane, where $x=\sin\varepsilon
 \cos\varphi$ and $y=\sin\varepsilon\sin\varphi$. The Cassini state C$_2$
 drifts along the $x$-axis during the integration (as indicated by the gray
 arrow), and reaches very large obliquity values.
 \label{fig4}}
\end{figure*}

Our results indicate that Jupiter spin axis could have been tilted either
when (i) the $s_8$ frequency swept over $\alpha_{\rm J}$ during the first
migration stage (that is before the instability happened), or when (ii)
the $s_7$ frequency approached $\alpha_{\rm J}$ at the end of planetary migration.
For (i) to work, the crossing of $s_8$ must be fast, such that the capture into
the resonance does not happen, but not too fast, such that some excitation
is generated by the resonance crossing. To obtain full $\varepsilon_{\rm J}=
3.1^\circ$ during this stage, the rate of change of $s_8$ during crossing,
${\rm d}s_8/{\rm d}t$, must be smaller than $0.05$ arcsec yr$^{-1}$~Myr$^{-1}$
(assuming that $I_{58}<0.05^\circ$). Since the evolution of $s_8$ mainly
relates to the radial migration of Neptune and dispersal of the outer disk
of planetesimals, this result implies that both these processes would need
to occur relatively slowly. More specifically, parameters ${\rm d}s_8/
{\rm d}t$ and $I_{58}$ would have to have values along a diagonal line
in the (${\rm d}s_8/{\rm d}t$, $I_{58}$) plane with larger values of $I_{58}$
requiring larger values of ${\rm d}s_8/{\rm d}t$ (see Fig.~\ref{fig3}).
Any model of planetary instability/migration can be tested against this
constraint. The models where ${\rm d}s_8/{\rm d}t$ is too slow and/or
$I_{58}$ is too large, as specified in Fig.~\ref{fig3}, can be ruled out,
because Jupiter's obliquity would be excited too much by the $s_8$ crossing.

\begin{figure}
\epsscale{0.8}
\plotone{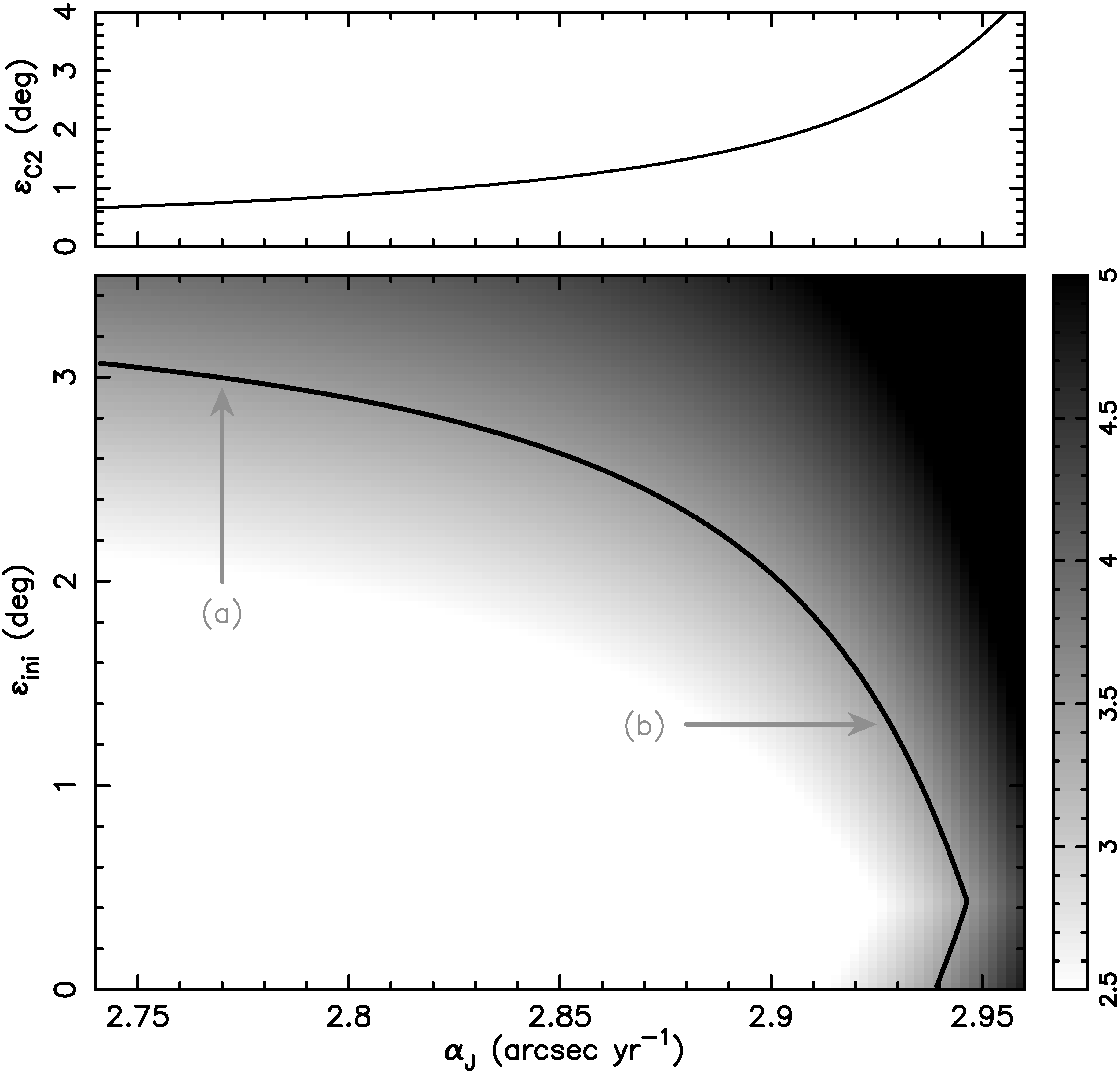}
\caption{Final obliquity of Jupiter resulting from an
 adiabatic approach of the $s_7$-frequency towards $\alpha_{\rm J}$
 (the obliquity has been computed in the reference frame
 associated with this frequency term in $\zeta$). While today's
 value of $s_7$ is known fairly accurately, $\alpha_{\rm J}$ has a
 significant uncertainty. We therefore treat $\alpha_{\rm J}$ as a
 free parameter. The initial obliquity of Jupiter is also treated as a
 free parameter, because it depends on the effects during the first
 migration stage (see Fig.~\ref{fig3}). The key to the shading
 scale is provided by the vertical bar on the right (white region
 corresponds to the final maximum obliquity smaller than $2.5^\circ$,
 incompatible with Jupiter's value). The bold curve corresponds to
 the $3.45^\circ$ isoline, estimated obliquity value today in this frame
 \cite[e.g.,][]{wc06}. The arrows indicate parametric location of the
 two examples shown on Fig.~\ref{fig6}. The upper panel shows the obliquity
 value of the Cassini state C$_2$ as a function of $\alpha_{\rm J}$.
 \label{fig5}}
\end{figure}

\begin{figure*}
\epsscale{1.0}
\plotone{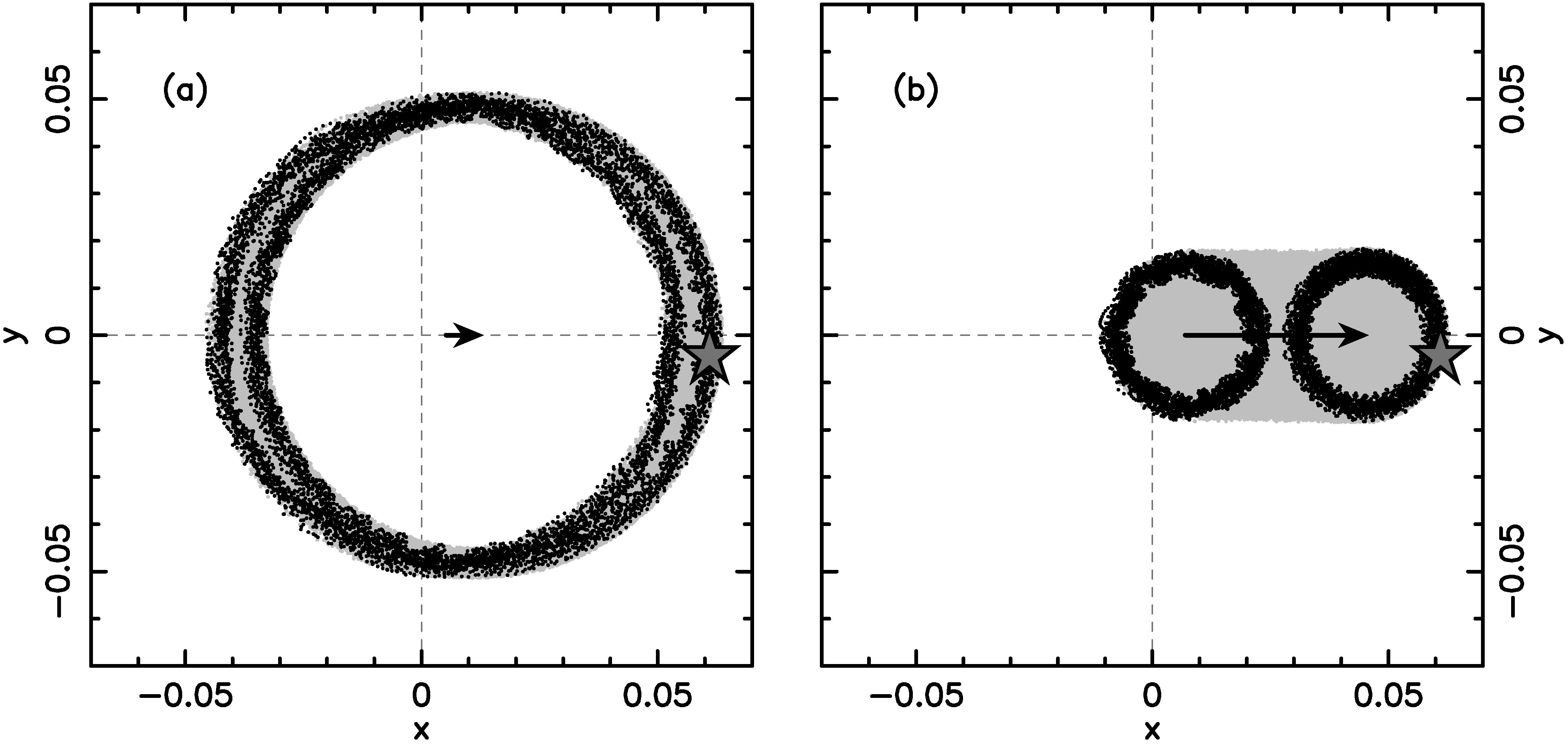}
\caption{Two examples of Jupiter's spin state evolution during
 the 1~Gyr-long time interval after the planetary instability.
 The planet pole is shown in the Cartesian coordinates $x=\sin
 \varepsilon\cos\varphi$ and $y=\sin\varepsilon\sin\varphi$.
 The arrow shows evolution of the Cassini state C$_2$ over the
 interval of time covered by the simulation. The gray star shows the
 current location of Jupiter's pole in these coordinates. Gray dots
 show the pole position output every $5$~kyr during the simulation,
 the black circles highlight the first and the last $30$~Myr of
 the evolution. Two different parametric combination along the
 bold curve in Fig.~\ref{fig5} were chosen: (i) $\alpha_{\rm J}=2.77$~arcsec
 yr$^{-1}$ and $\varepsilon_{\rm ini}=3.1^\circ$ in the left panel (a), and
 (ii) $\alpha_{\rm J}=2.93$~arcsec yr$^{-1}$ and $\varepsilon_{\rm ini}
 =1.3^\circ$ in the right panel (b).
 \label{fig6}}
\end{figure*}

Not much excitation of $\varepsilon_{\rm J}$ is expected during the $s_8$
crossing if ${\rm d}s_8/{\rm d}t$ was relatively fast and/or if $I_{58}$ was
only a very small fraction of its current value. If that is the case, Jupiter's
obliquity would probably need to be excited during the very last stages of
migration by (ii). For that to work, Jupiter's precession constant $\alpha_{\rm J}$
would need to be $\simeq 2.95$~arcsec yr$^{-1}$ (assuming $\varepsilon_{\rm J}=0$
initially), which is a value that is significantly larger than the one estimated
by \cite{hetal11}. This means that Helled's model would need to be adjusted
to fit within this picture. It is also possible, however, that Jupiter's
present obliquity was contributed partly by (i) and partly by (ii). If so,
Figs.~\ref{fig3} and \ref{fig5} express the joint constraint on ${\rm d}s_8/
{\rm d}t$ and $I_{58}$ during the first stage, and $\alpha_{\rm J}$.

As for Saturn, our results indicate that the capture into the spin-orbit
resonance with $s_8$ \citep{wh04,hw04} is indeed possible during the late
stages of planetary migration, assuming that the migration rate was slow
enough. The exact constraint on the slowness of migration depends on $I_{68}$,
which in turn depends how much Neptune's inclination was excited by the
instability and how long it remained elevated \citep{betal09}. Since in the
NM12 models, Neptune's orbital inclination is never large, we have good
reasons to believe that $I_{68}$ was comparable to its current value when the
crossing of $s_8$ occurred. Thus, using $I_{68}\simeq 0.064^\circ$, we find
that the e-folding migration timescale $\tau$ would need to be $\tau \gtrsim
100$~Myr. If $\tau > 200$~Myr, however, the capture in the $s_8$ resonance
would be strictly adiabatic. This would imply, if $\varepsilon_{\rm S}$ was
negligible before capture, that the resonant state should have a very small
libration amplitude \cite[see][]{wh04,hw04}. It would then be difficult to
explain the current
orientation of Saturn's spin axis, which indicates that the libration
amplitude should be at least $30^\circ$. A more satisfactory solution,
however, can be obtained for $\tau \simeq 100-150$~Myr, in which case the
capture into the resonance was not strictly adiabatic. In this case the
$\geq 30^\circ$ libration amplitude is obtained during capture.

\begin{figure}
\epsscale{0.8}
\plotone{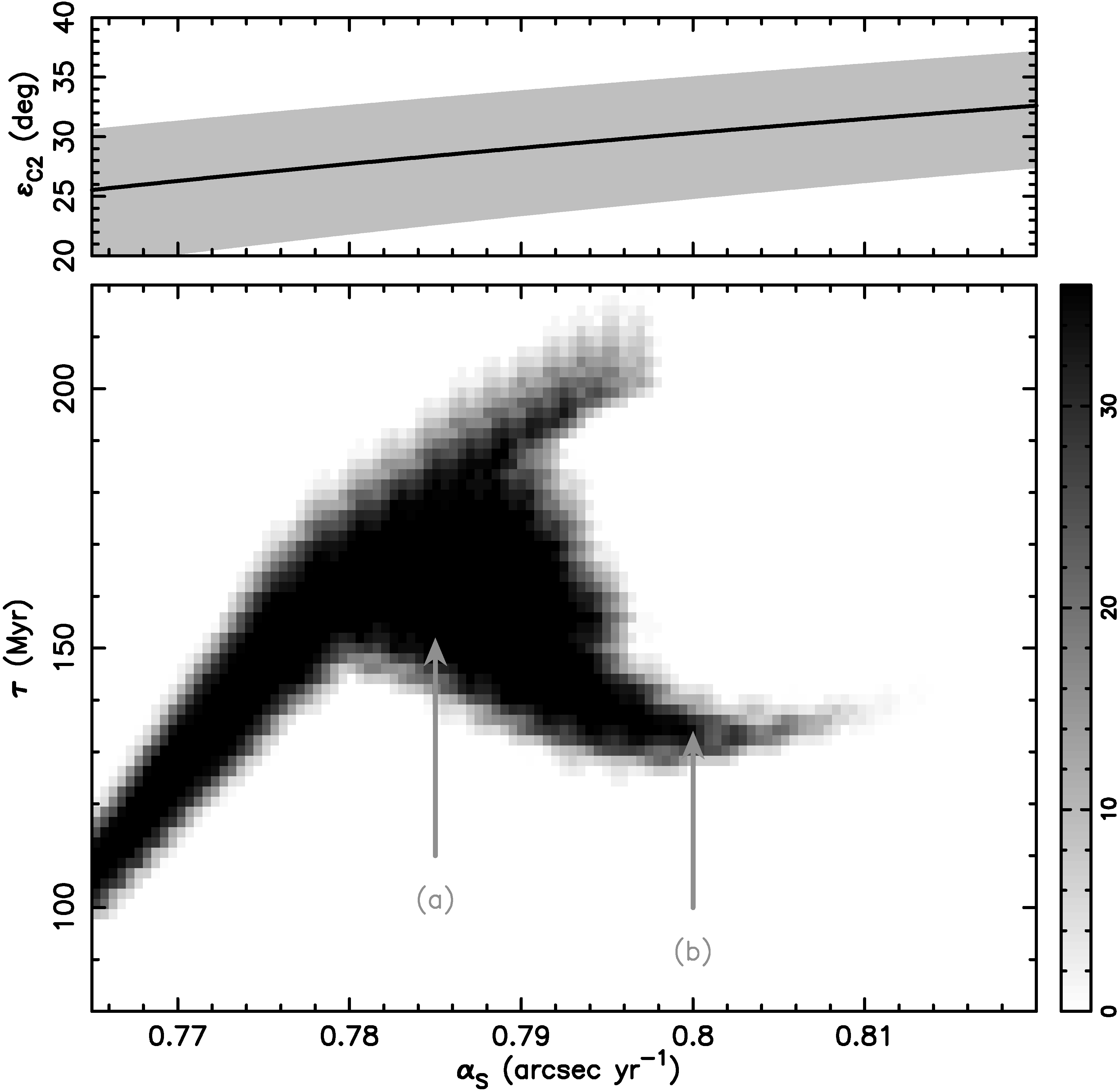}
\caption{Bottom panel: Distribution of solutions successfully matching
 Saturn's spin state in the parametric plane $\alpha_{\rm S}$ (abscissa) vs $\tau$
 (ordinate), the e-folding time of $s_8$-frequency evolution during the phase~2.
 No successful solutions were obtained in the white region of the plot.
 The darker the gray-scale in the given bin, the more robust the solution
 is. The maximum value 36 corresponds to 36 sampled initial conditions
 of the simulations for each $(\alpha_{\rm S},\tau)$ pair (see the side bar).
 When $\alpha_{\rm S}
 \simeq 0.78$ arcsec~yr$^{-1}$, the obliquity of the corresponding to Cassini
 state C$_2$ is $\simeq 27.4^\circ$. Therefore the capture solution corresponds
 to the minimum needed libration amplitude of Saturn's spin in the $s_8$
 reference frame. From this value the larger libration-amplitude solutions
 bifurcate towards smaller/larger $\alpha_{\rm S}$ values. The arrows indicate
 parametric location of two particular examples shown on panels (a) and (b)
 of Fig.~\ref{fig8}.
 Top panel: Obliquity of the Cassini state C$_2$ (solid line) and maximum
 width of the associated resonant zone (gray area) as a function of
 $\alpha_{\rm S}$. We assume inclination $I=I_{68}=0.064^\circ$ and
 terminal value of the orbital precession frequency $s_8\simeq -0.692$
 arcsec~yr$^{-1}$.
 \label{fig7}}
\end{figure}

\begin{figure}
\epsscale{1.0}
\plotone{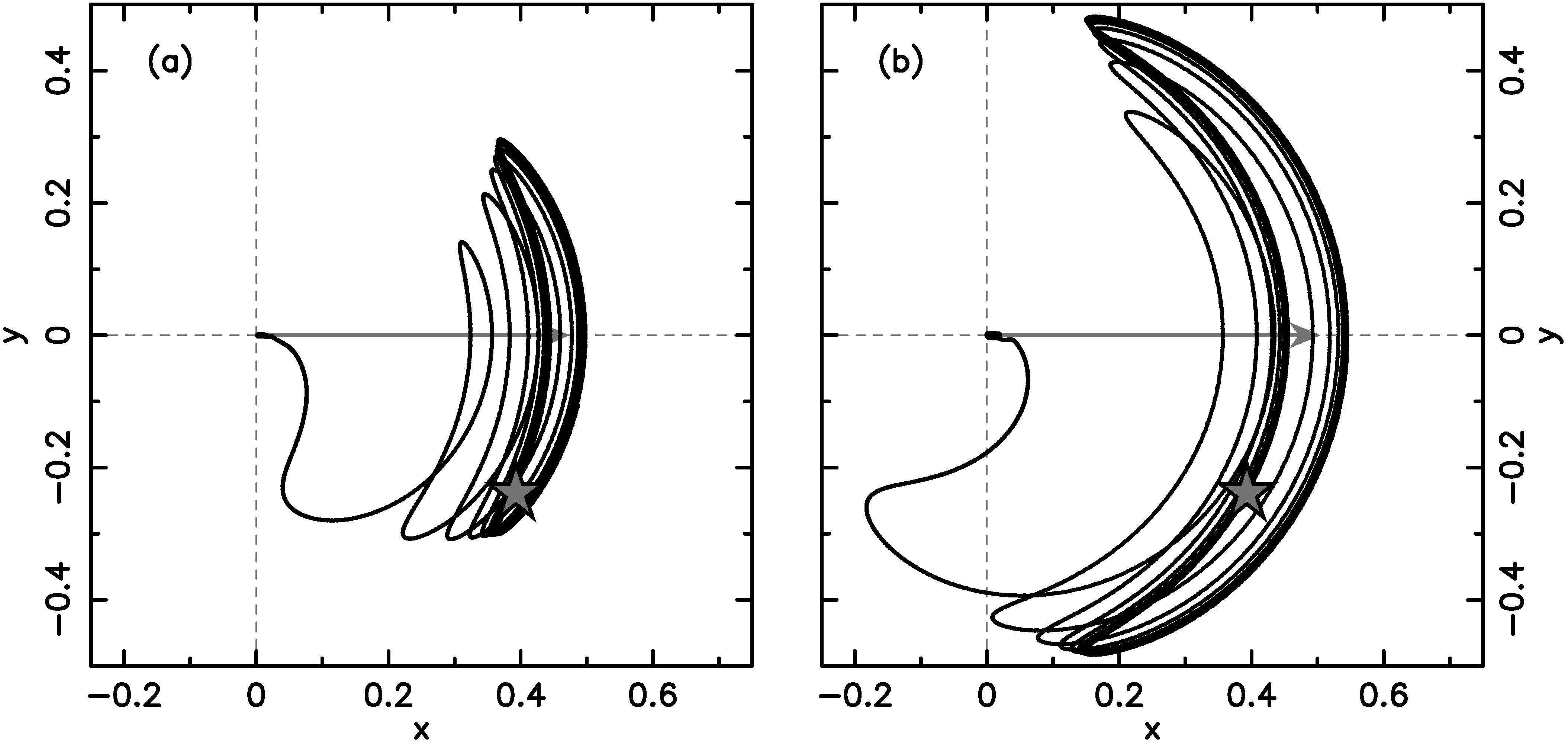}
\caption{Two examples Saturn's obliquity excitation by the capture and
 evolution in the resonance with $s_8$. The spin axis ${\bf s}$ is
 projected onto the $(x,y)$ plane, where $x=\sin\varepsilon\cos\varphi$
 and $y=\sin\varepsilon\sin\varphi$. The reference frame used here is
 the one associated with the $s_8$ frequency term contributing to $\zeta$
 of Saturn. The gray arrow shows the evolution of the Cassini state C$_2$ over
 the full length of the simulation ($1$~Gyr). In panel (a), we assumed that
 $\alpha_{\rm S}=0.785$~arcsec yr$^{-1}$, which means that the terminal
 obliquity of C$_2$ is at $\simeq 28.2^\circ$, and $\tau=150$ Myr. In (b),
 we used $\alpha_{\rm S}=0.8$~arcsec~yr$^{-1}$, which means that the terminal
 obliquity of C$_2$ is at $\simeq 30.3^\circ$, and $\tau=135$~My. These
 combinations were also highlighted by arrows on Fig.~\ref{fig7}. The star
 in each panel shows the present orientation of the Saturn pole vector in
 the $(x,y)$ coordinates \cite[e.g.,][Table~2]{wh04}.
\label{fig8}}
\end{figure}

While the capture conditions pose a strong constraint on the timescale of
Neptune's radial migration, as discussed above, and additional constraint
on Saturn's precession constant $\alpha_{\rm S}$ derives from the present
obliquity of Saturn. This is because, again assuming that Saturn spin
vector is in the resonance with $s_8$ today, the present obliquity of Saturn
implies that the Cassini state C$_2$ of this resonance would have to be located at
$\varepsilon\sim (28-30)^\circ$. This would require that $\alpha_{\rm S}\simeq
0.78-0.80$ arcsec~yr$^{-1}$. The value derived by \cite{hetal09} is larger,
$\simeq 0.8445$ arcsec~yr$^{-1}$, and clearly incompatible with this assumption.
Direct measurements of the mean precession rate of Saturn's spin axis suggest
that $\alpha_S\simeq 0.81$ arcsec~yr$^{-1}$, which is still slightly larger
than the range given above, but the uncertainty interval of this estimate
includes values below $0.8$ arcsec~yr$^{-1}$ (R.~A. Jacobson, personal
communication). Figuring out the exact value of Saturn's precession constant
will therefore be important. Once $\alpha_{\rm S}$ is known, Fig.~\ref{fig7}
could be used to precisely constrain the timescale of planetary migration.

\acknowledgments
 We are grateful to Robert Jacobson for sharing his latest estimate of the
 precession rate of Saturn's spin axis. We also thank Tristan Guillot and
 Alessandro Morbidelli for discussions and pointing to us the future capability
 of Juno mission to constrain Jupiter's precession constant. The work of DV 
 was supported by the Czech Science Foundation (grant GA13-01308S). The work
 of DN was supported by NASA's Origin of Solar Systems (OSS) program.


\clearpage

\clearpage

\clearpage

\clearpage

\clearpage

\clearpage

\clearpage

\end{document}